\documentclass[11pt,psfig]{article}
\newcommand{\Ltwo}{$\mathcal{L}_{2}~$}

\usepackage{times}
\usepackage{amssymb}
\usepackage{amsmath}
\usepackage{graphicx}
\usepackage[letterpaper,total={6.5in, 9in}, top=1.0in, bottom=1.0in, left =0.5in, right=1.5in,marginparwidth=1.0in]{geometry}

  \renewenvironment{thebibliography}[1]{%
    \begin{oldthebibliography}{#1}%
      \setlength{\parskip}{0ex}%
      \setlength{\itemsep}{0.14ex}%
  }%
  {%
    \end{oldthebibliography}%
  }

\begin{document}

\newenvironment{part1}
{\begin{minipage}[t] 
  {0.45 \textwidth}}
  {\end{minipage}}
\newenvironment{part2}
{\hfill\begin{minipage}[t] 
  {0.45 \textwidth}}
  {\end{minipage}}
  

\title{{\bf Zames-Falb Multipliers: don't panic}}
\author{\bf Matthew C. Turner  \\
             Cyber Physical Systems Research Group, \\
             Electronics and Computer Science, \\
             University of Southampton, \\
             Southampton, \\
             SO17 1BJ,
             UK. \\ Email:  {\tt m.c.turner@soton.ac.uk}
             }

\date{}

\maketitle 


\begin{abstract}
Zames-Falb multipliers are mathematical constructs which can be used to prove stability of so-called Lur'e systems: systems that consist of a feedback interconnection of a linear element and a static nonlinear element. The main advantage of Zames-Falb multipliers is that they enable ``passivity''-like results to be obtained but with a level of conservatism much lower than \emph{pure} passivity results. However, some of the papers describing the development of the Zames-Falb multiplier machinery are somewhat abstruse and not entirely clear. This article attempts to provide a relatively simple construction of Zames and Falb's main results which will hopefully be understandable to most graduate-level control engineers.
\end{abstract}


\section{Introduction}

%
 \begin{figure}[!h]
   \includegraphics[width=0.5\textwidth]{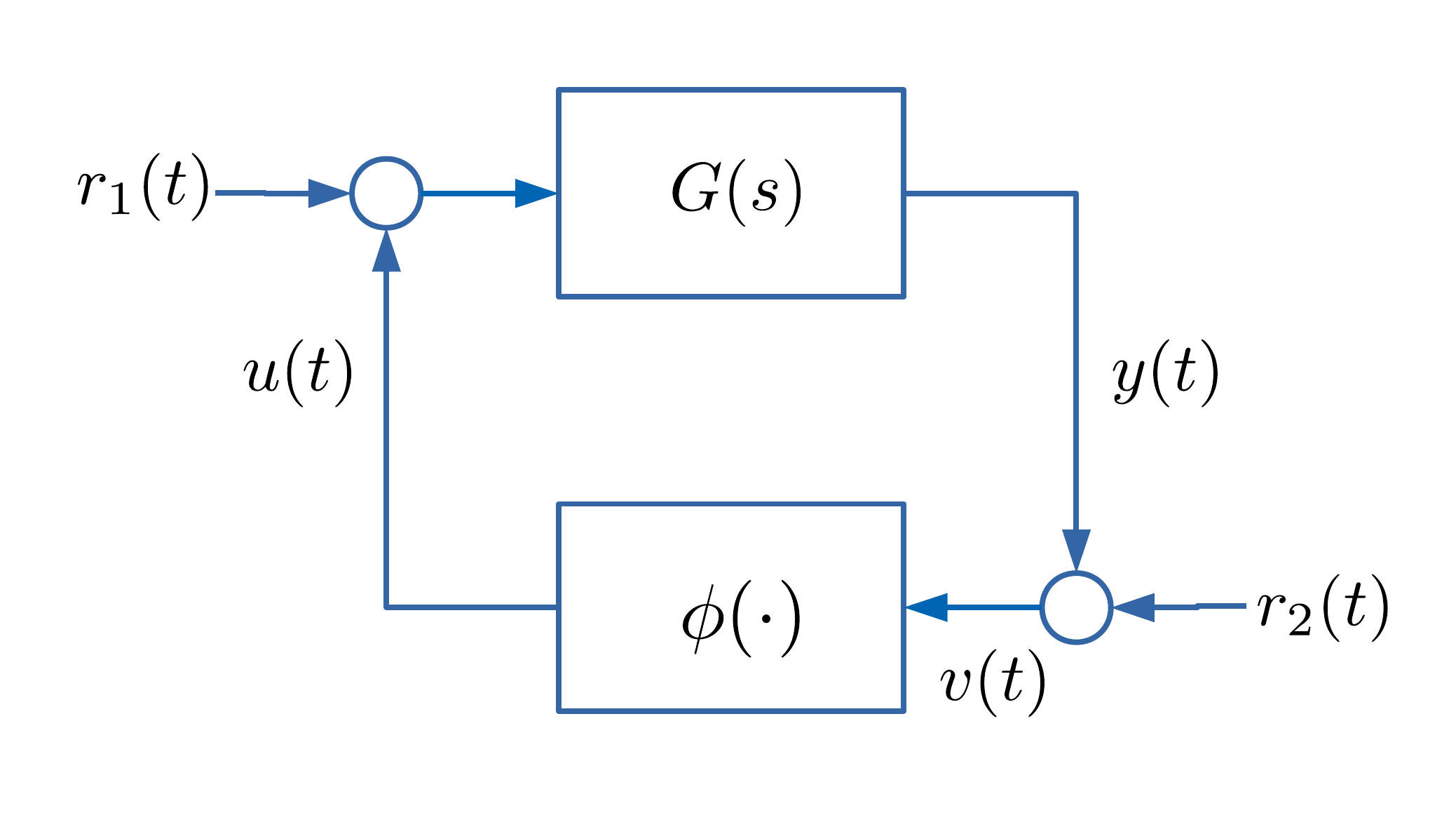}
   \parbox{0.5\textwidth}{ \vspace{-2in}
   \caption{\label{fig:Pfig5} 
   Lur'e system: a feedback interconnection of a linear system and a static nonlinearity. In this article positive feedback is assumed.
   }}
 \end{figure}

Absolute stability theory is the branch of control theory concerned with establishing stability of a feedback interconnection of the type depicted in Figure 1. It is routinely taught in graduate courses on nonlinear control, but often only the popular techniques of the Circle and Popov Criteria are covered, both of which assume the nonlinearity satisfies so-called sector bounds. This is partly because the Circle and Popov Criteria have a nice Lyapunov interpretation \cite{Khalil96} and partly because they have an appealing graphical form (in the single-input-single-output case). They can however be quite conservative, sometimes not providing stability guarantees when the feedback interconnection is, in fact, stable. 

When one is looking for less conservative results than the Circle and Popov Criteria, one's attention is inevitably drawn to the so-called ``Zames-Falb'' multipliers which, in a sense,  generalise the Circle and Popov Criteria to the case when the nonlinearity is \emph{slope-restricted}: a tighter condition than sector-boundedness. Numerous papers \cite{TurnerIEEETAC2012,CarrascoSCL2014,SchererIJRNC2018} have demonstrated the superiority of the Zames-Falb multipliers over the Circle/Popov Criteria in terms of lower levels of conservatism. However, a treatment of Zames-Falb multipliers remains quite difficult to incorporate into a typical UK-style Master's course. 

This article is an attempt to address this problem and has been initiated by student questions on further reading in my graduate-level Nonlinear Control course, and also by questions I have been asked since I published the articles \cite{TurnerIEEETAC2009,TurnerIJRNC2012} which were based on Zames-Falb multipliers. Central to many of these questions were some fundamental issues with the understanding of Zames-Falb multipliers. While there are several excellent texts treating Zames-Falb multipliers and related concepts (\cite{Vidyasagar:book,Willems:book}) and several papers which also attempt to provide more background for the reader (\cite{HeathCDC2005}), most of these texts are somewhat terse and difficult for the typical graduate control engineer to understand. In addition, the term ``Zames-Falb Multiplier'' seems to be rather off-putting and a cause of anxiety in the student of absolute stability. My aim here is to provide a fairly simple and easy to follow introduction to Zames-Falb multipliers which prioritises clarity over succinctness; hence the title. I hope this is of use to readers and fills a niche in the literature on Zames-Falb multipliers. 

\subsection{Scope of the article}

This article is entirely different from the review paper \cite{CarrascoEJC2016} which provides more of an overview of the state-of-the-art, and really concentrates on searches and properties of Zames-Falb multipliers. Instead, this article attempts to give a complete but understandable proof of the main results of Zames-Falb, introduced in their paper \cite{ZamesSIAM1968}. The article also tries to give some background to the work of Zames-Falb and towards the end touches on the problems associated with \emph{actually finding} or \emph{computing} Zames-Falb multipliers - this is biased towards my own research. 

This article is \emph{not} a review of absolute stability in general, nor does it cover Zames' other significant contribution on conic sector stability analysis \cite{ZamesIEEETAC1966a,ZamesIEEETAC1966b} (see also \cite{SafonovIEEETAC1981,BridgemanIJC2014,TurnerIJRNC2020}).   

\subsection{Style of article}

\marginpar{\emph{``I like the cover,'' he said. ``Don't Panic. It's the first helpful or intelligible thing anybody's said to me all day.''}}

I have tried to keep the article as readable as possible but inevitably there is a certain amount of notation and some non-trivial concepts which need introducing in order to establish the results. Most of the notation follows that in \cite{Zhou96}. The article tries to use simple terminology and language where possible and the treatment of some concepts (such as causality and well-posedness) are deliberately light-touch. There is a certain amount of whimsicality in the writing which is there to, hopefully, ensure the article is fairly light. The quotes from Reference \cite{Adams:book} are hopefully appropriate and in keeping with the nature of the article.

\section{Systems under consideration}

Zames-Falb multipliers were developed to tackle the absolute stability problem. The best way to understand the absolute stability problem is from the diagram in Figure \ref{fig:Pfig5}, which depicts a feedback interconnection containing a linear-time-invariant (LTI) system $\mathbf{G}$ and a static feedback nonlinearity $\phi (\cdot)$ - this interconnection is sometimes called a Lur'e/Lurie/Lurye system. The linear system $\mathbf{G}$ can be thought of as a linear operator, with an associated transfer function $G(s)$ and impulse response $g(t)$. It is also possible to interpret it as a state-space system, although this is not required here.  

The nonlinearity $\phi(.)$ here is considered to be \emph{static} (sometimes also called memoryless, it has no dynamics) and \emph{scalar valued}, that is
\[
 \phi(\cdot): \mathbb{R} \mapsto \mathbb{R}
\]
The absolute stability problem involves finding conditions which guarantee stability of the interconnection for \emph{classes} of nonlinearity i.e. we seek stability for all $\phi \in \mathcal{N}$ where $\mathcal{N}$ is some class of nonlinearity. In the Circle and Popov Criteria the class of nonlinearities $\mathcal{N}$ is the class of sector bounded nonlinearities, but different classes - introduced shortly - will be of interest as well.

\textbf{Remark -- Robust Control.} The absolute stability problem is, in essence, a form of \emph{robust control} problem where the nonlinearity $\phi(\cdot)$ represents the \emph{uncertainty} in the system. In the robust control literature, $\phi (\cdot)$ is often labelled $\Delta$ where $\Delta$ represents a broader class of uncertainty - see, for example, \cite{MegretskiIEEETAC1997}. \hfill $\square$

\subsection{Stability}

In this article we are concerned with \emph{input-output} stability and more specifically with so-called \Ltwo stability. Central to this definition is the so-called \Ltwo space and its extension
\begin{align}
 \mathcal{L}_2 & := \left\{ x (t) ~ : ~ \int_{0}^{\infty} \| x(t) \|^2 ~ dt < \infty \right\} \\
 \mathcal{L}_{2e} & := \left\{ x(t) ~ : ~ \int_0^T \| x(t) \|^2 ~ dt < \infty \quad \forall \quad  0 \leq T < \infty \right\} 
\end{align}
The space \Ltwo represents all \emph{square integrable} signals and roughly speaking represents most \emph{well behaved signals}. The space $\mathcal{L}_{2e}$ represents most signals which we may have to deal with and it is important since, from an operator perspective we represent systems as mappings between $\mathcal{L}_{2e}$ spaces. For example an LTI system $\mathbf{G}$ can be considered as a mapping between these spaces, that is
\[
 \mathbf{G}: \mathcal{L}_{2e} \rightarrow \mathcal{L}_{2e}
\]
From an operator perspective, the nonlinearity $\phi$ also maps $\mathcal{L}_{2e}$ to $\mathcal{L}_{2e}$. 

With reference to Figure \ref{fig:Pfig5} stability is then defined with respect to the signal space $\mathcal{L}_2$. Specifically, given a LTI system $\mathbf{G}$ and a static nonlinearity $\phi (\cdot)$:

\begin{center}
\framebox[0.9\textwidth]{~ 
\parbox{0.85\textwidth}{ \vspace{0.1in} The feedback system in Figure \ref{fig:Pfig5} is said to be \emph{stable} if the signals $u(t),y(t) \in \mathcal{L}_2$ for all exogenous signals $r_1(t),r_2(t) \in \mathcal{L}_2$. 
 \vspace{0.1in}}  }
\end{center}

In other words:

\centerline{\emph{``Well behaved inputs lead to well behaved outputs''}}

Good references on \Ltwo stability are \cite{Vidyasagar:book, Limebeer:book} and \cite{Jonsson:notes}. Here it is sufficient to understand stability as defined above. 

A standing assumption throughout the paper is that the LTI element in Figure \ref{fig:Pfig5} is stable and causal: its transfer function contains only open-left-half-plane poles. Such a system is (again following the robust control literature) labelled $\mathcal{RH}_{\infty}$. Specifically, the assumption
\[
 \mathbf{G} \in \mathcal{RH}_{\infty}
\]
is made throughout the paper. The assumption made on the Zames-Falb multiplier, introduced shortly, will not be as strict. In this case, the only requirement is that the multiplier is bounded on the imaginary axis\footnote{Here, the multiplier is assumed to be rational, although this is not actually \emph{required}.}. This is often denoted as
\[
 \mathbf{M} \in \mathcal{RL}_{\infty}
\]
and can be interpreted as $\mathbf{M}$ consisting of both causal and anti-causal elements:
\[
 \mathbf{M} = \mathbf{M}_c + \mathbf{M}_a 
\]
where $\mathbf{M}_c \in \mathcal{RH}_{\infty}$ and $\mathbf{M}_a \in \mathcal{RH}_{\infty}^{\perp}$ i.e. $\mathbf{M}_a$ contains poles only in the open right half complex plane. For much of this article, one may essentially ignore this detail, but it is stressed that for searches there is often a useful advantage in allowing the multiplier to have an anti-causal component \cite{CarrascoSCL2014}. On a first read this subtlety can probably be ignored.

\marginpar{\small \emph{``Just believe everything I tell you, and it will all be very, very simple.'' \\ ``Ah, well, I'm not sure I believe that''}}

\subsection{Nonlinearities} 

\begin{figure}[!ht]
   \includegraphics[width=0.34\textwidth]{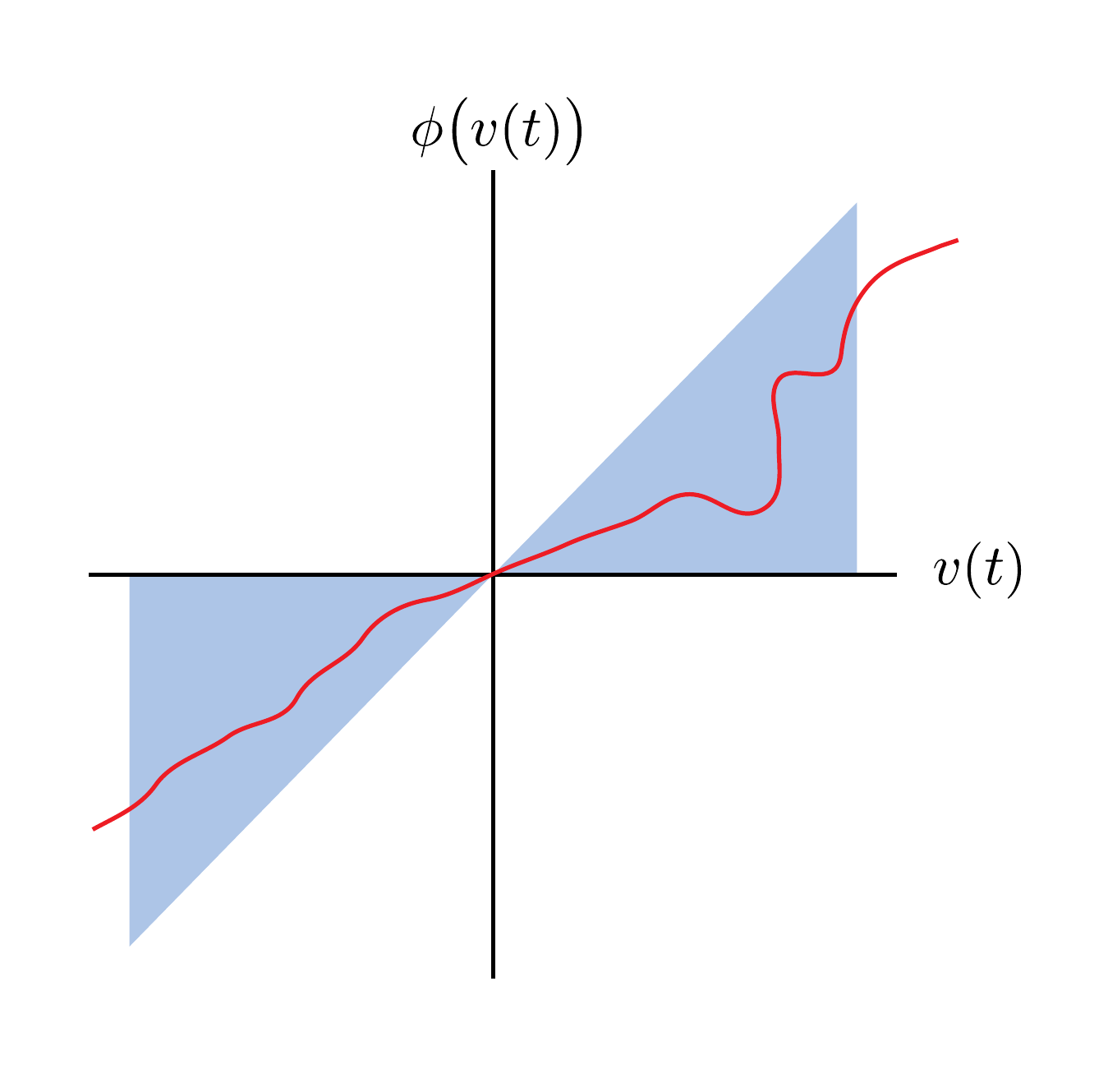}~
  \includegraphics[width=0.34\textwidth]{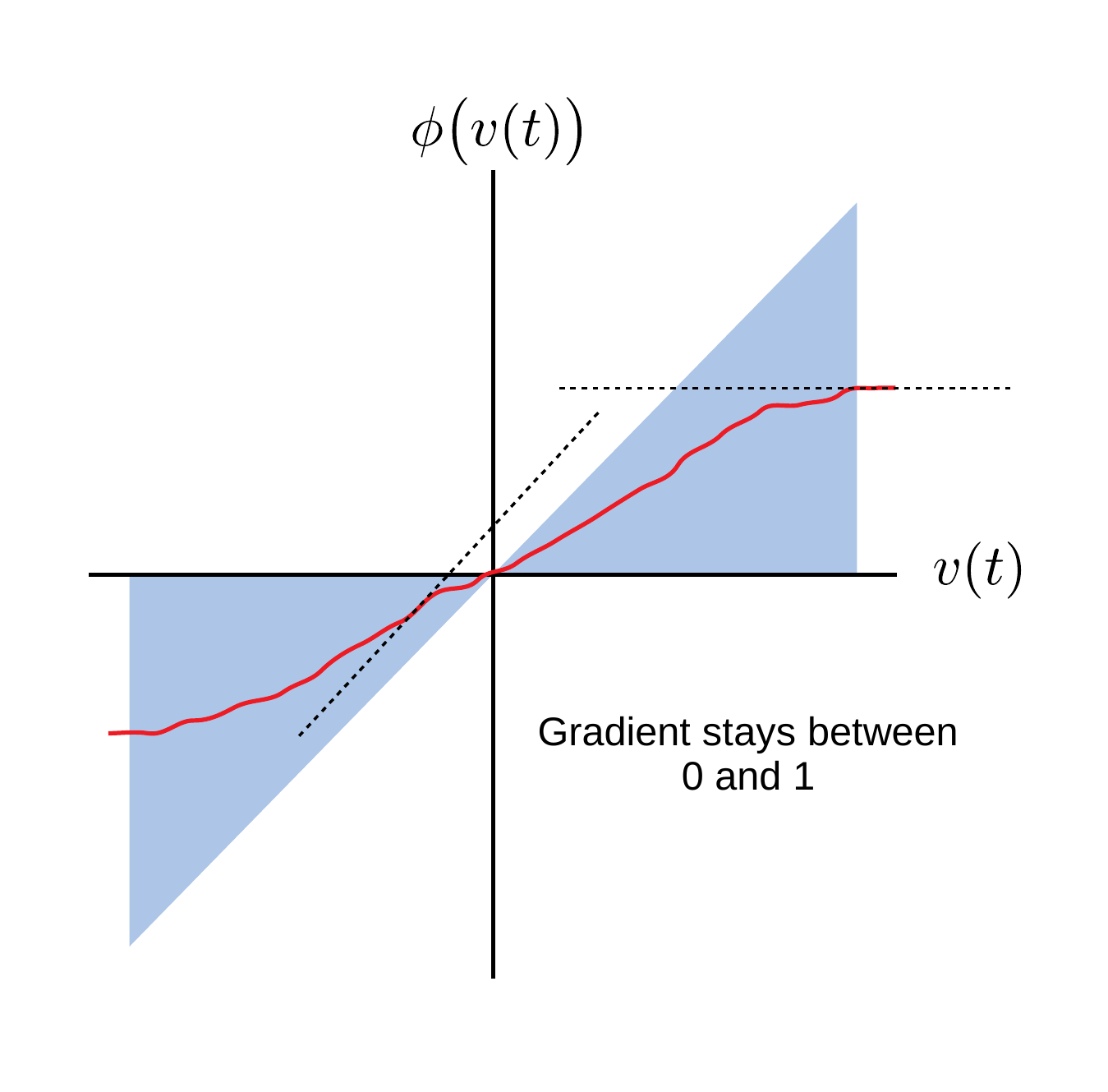}
   \parbox{0.3\textwidth}{ \vspace{-2in}
   \caption{\label{fig:non} 
   Nonlinearities: the left-hand diagram shows sector $[0,1]$ and an example nonlinearity; the right hand diagram shows an example slope-restricted nonlinearity.
   }}
 \end{figure}

The classical Circle and Popov criteria both consider so-called sector bounded nonlinearities in which the nonlinearity $\phi(.)$ satisfies the \emph{sector bounds}
\begin{align}
 \alpha_1 \leq \frac{\phi(\sigma)}{\sigma} \leq \alpha_2 \quad \forall \sigma \in \mathbb{R}
 \label{eq:sec}
\end{align}
where $\alpha_1 < \alpha_2$ are two real, not necessarily positive constants. The short hand
\[
 \phi \in {\rm Sector}[\alpha_1,\alpha_2]
\]
is used to indicate a nonlinearity satisfying inequality (\ref{eq:sec}). 
Many nonlinearities satisfy sector bounds: the saturation and deadzone functions being two common ones. The concept of a sector is illustrated in Figure \ref{fig:non}; note often the lower sector bound is taken as zero (something which can be achieved with a loopshift - see \cite{Khalil96}, Chapter 10). 

The main problem with the sector bounds is that many nonlinearities inhabit the same sector and thus results obtained using sector information tend to be quite conservative because they need to hold for \emph{all} nonlinearities in the same sector. To overcome this, various researchers (\cite{OShea66,ZamesSIAM1968,ParkIEEETAC2002}) have used the concept of \emph{slope restrictions} to lessen the conservatism of the results. A slope restricted nonlinearity is one where the generalised derivative of the nonlinearity is assumed to lie between certain bounds. Denoting $\partial \phi(.)$ as the generalised derivative of $\phi(.)$, the slope restriction is equivalent to requiring
\[
 \partial \phi \in [0,\alpha]
\]
which is often written as
\begin{align}
 0 \leq \frac{\phi(x)-\phi(y)}{x-y} \leq \alpha \quad \forall x,y  \in \mathbb{R} 
\end{align}
where the lower bound on the slope has been taken to be zero (again possibly by a loopshift). This therefore implies (setting $y=0$ and under the assumption that $\phi(0)=0$) that $\phi \in {\rm Sector}[0,\alpha]$, but the converse is not generally true. 

Two final classes of nonlinearities are needed in order to arrive at the Zames-Falb results: the classes of \emph{monotone} and \emph{odd monotone} nonlinearites. These classes of nonlinearities are essentially 2nd-4th quadrant (odd) nonlinearities with a linear bound on their ``gain''. They are of independent interest, but are also the stepping stone to perhaps the most useful Zames-Falb results: those devoted to slope-restricted nonlinearities. Formally, these classes of nonlinearities are defined below.

\newtheorem{defn}{Definition}
\begin{defn}
A nonlinearity $\phi(\cdot)$ is said to be a \emph{monotone non-decreasing} nonlinearity if
\[
 \Big( \phi(x)-\phi(y) \Big) (x-y) \geq 0 \quad \forall x,y \in \mathbb{R}
\]
 and it is bounded $| \phi(x) | \leq \gamma | x |$ for some $\gamma >0$ and all $x \in \mathcal{L}_2$.
A nonlinearity satisfying these properties is said to belong to $\mathcal{N}_{M}$. In addition if $\phi (.)$ is odd, that is $\phi(x) = -\phi(-x)$, then the nonlinearity is said to belong to $\mathcal{N}_{MO}$. 
\end{defn}

Although, strictly speaking, the above definition defines monotone \emph{non-decreasing} nonlinearities, with some abuse of terminology, they are often described as \emph{monotone increasing} nonlinearities: in this article no distinction will be made.

\subsection{A few further concepts and notation}

A few further concepts need emphasizing or introducing. In addition to the plant $\mathbf{G}$, it is often necessary to introduce other linear operators which will be denoted in bold font. A linear operator $\mathbf{P}$ will have an associated Laplace transform $P(s)$, Fourier transform $P(j \omega)$ and impulse response $p(t)$. It is recalled that the output (in the time-domain) of such a system subject to an input $u(t)$ may be calculated using the convolution integral
\[
 y(t) = \int_{-\infty}^{\infty} g(\tau) u(t-\tau) d \tau
\]
When $\mathbf{G}$ is causal and the output at time $t=0$ is zero, this can be replaced by 
\[
 y(t) =  \int_{0}^{t} g(\tau) u(t-\tau) d \tau
\]
The Fourier transform of a signal $u(t)$, denoted $\hat{u}(j \omega)$, is needed for the IQC approach introduced shortly. 

Stability analysis based on input-output notions of passivity is intrinsically linked to the concept of an inner product and this is also used frequently in the IQC approach. The inner product of two signals $g(t)$ and $h(t)$ is defined as 
\[
 \langle g(t),h(t) \rangle = \int_{-\infty}^{\infty} g(t) h(t) ~ dt
\]
By Parseval's Identity (e.g. Appendix B in \cite{Vidyasagar:book}), this is equivalent to the frequency domain inner product
\[
\langle \hat{g}(j \omega),\hat{h}( j \omega) \rangle = \frac{1}{2 \pi} \int_{-\infty}^{\infty} \hat{g}^*(j \omega) \hat{h}(j \omega) ~ d \omega
\]
where $(\cdot)^*$ represents complex conjugate transpose.  Obviously if the time-domain inner product is non-negative, then so is the frequency domain inequality. 

The main technical idea of the Zames-Falb approach is to ensure a certain integral is non-negative. It transpires that this is the case if the so-called $\mathcal{L}_1$ norm of a certain ``system'' (linear operator) is less than unity. Given a linear operator $\mathbf{P}$, the $\mathcal{L}_1$ norm is defined as
\[
 \| \mathbf{P} \|_1 = \int_{-\infty}^{\infty} |p(t)| dt 
\]
In the subsequent sections it will be seen that this norm arises naturally in the analysis.

\subsection{Multipliers}

 \begin{figure}[!h]
   \includegraphics[width=0.6\textwidth]{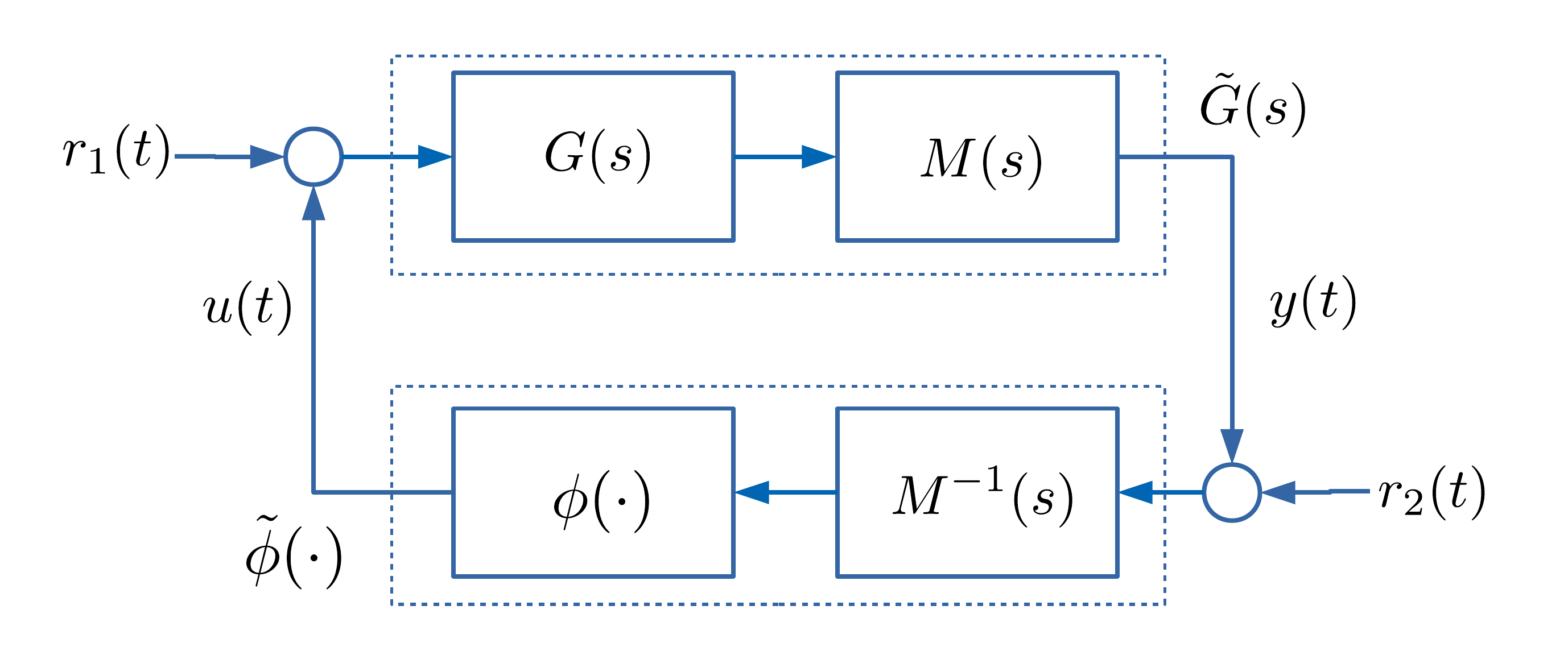}
   \parbox{0.4\textwidth}{ \vspace{-2in}
   \caption{\label{fig:Pfig_multiplier} 
   Lur'e system with multipliers: Objects $\mathbf{M}$, called \emph{multipliers}, assist with stability assessment
   }}
 \end{figure}
 
 \marginpar{\small \emph{In those days spirits were brave, the stakes were high, men were real men, women were real women and small furry creatures from Alpha Centauri were real small furry creatures from Alpha Centauri.}}

There are various approaches one might take to prove stability of the system in Figure \ref{fig:Pfig5}. In the 1960s and 1970s the favoured approach was to use \emph{multipliers}, which is where the term \emph{Zames-Falb multiplier} originates. The idea with multipliers effectively comes from the desire to overcome limitations imposed by \emph{passivity}. Referring again to Figure \ref{fig:Pfig5}, it is well known that the (negative) feedback interconnection is stable if $\mathbf{G}$ and $\phi(\cdot)$ are both passive: actually it is more subtle, $\mathbf{G}$ must be strictly passive, but a discussion of this is out of scope. The key point to note is that passivity places a \textbf{very strong requirement on $\mathbf{G}$} which is often not, and in practice rarely (never?), satisfied. To overcome this strict requirement, one can introduce a so-called multiplier $\mathbf{M}$, whose inverse $\mathbf{M}^{-1}$ is well-defined, and re-draw Figure \ref{fig:Pfig5} as in Figure \ref{fig:Pfig_multiplier}. Using the same passivity arguments it follows that the  interconnection is stable if the LTI system $\mathbf{\tilde{G}}=\mathbf{MG}$ is strictly passive and the new nonlinearity $\tilde{\phi} = \phi \circ \mathbf{M}^{-1}$ is passive. The advantage of this new arrangement is that by introducing the multiplier $\mathbf{M}$ the passivity requirements are on $\mathbf{GM}$ and are hopefully easier to enforce. 

\textbf{Remark: informality.} This section has been written in an informal manner. These concepts are used extensively in classical texts describing Zames-Falb multipliers \cite{Willems:book,Vidyasagar:book} as well as in the original papers on the subject. However, the approach taken in this article differs somewhat to that described above. \hfill $\square$

\subsection{Integral quadratic constraints (IQC's)}

Instead of using multipliers directly, the proofs given here are based on the IQC approach to stability analysis \cite{MegretskiIEEETAC1997}. The IQC approach is powerful because it a) captures existing results in robust control; and b) enables one to combine different stability results into a common analysis framework: The Zames-Falb multiplier results are a good fit to this. A further advantage of the IQC results is that the approach is arguably more direct than the multiplier approach and dispenses with some of the loopshifting involved.

The key to applying IQC results is to establish that a nonlinearity satisfies an IQC. In the frequency domain the nonlinearity $\phi(\cdot)$ is said to \emph{satisfy the IQC defined by $\Pi (j \omega)$}, where $\Pi (j \omega) = \Pi^* (j \omega)$ and ${\Pi} \in \mathcal{RL}_{\infty}$, if the following inequality holds for all $\hat{v}(j \omega)$:
\begin{align}
 \int_{-\infty}^{\infty}
 \begin{bmatrix}
  \hat{v} (j \omega) \\ \widehat{\phi(v (j \omega))}
 \end{bmatrix}^*
 \Pi (j \omega) 
 \begin{bmatrix}
  \hat{v} (j \omega) \\ \widehat{\phi(v (j \omega))}
 \end{bmatrix} ~ d \omega \geq 0 
 \label{eq:IQCgen}
\end{align}
Such an IQC is often established, and will be in this article, by first deriving a time-domain integral quadratic constraint and then converting this into the frequency domain form (\ref{eq:IQCgen}) via Parseval's Identity. 

The beauty of establishing an IQC of the form (\ref{eq:IQCgen}) is that the IQC theory established in \cite{MegretskiIEEETAC1997} can then be directly applied in order to obtain a frequency domain stability condition, as indicated in the theorem below. 

\newtheorem{theorem}{Theorem}
\begin{theorem}[\cite{MegretskiIEEETAC1997}]
 Let $\mathbf{G} \in \mathcal{RH}_{\infty}$ and assume $\phi(\cdot)$ is a bounded, causal operator. Assume that:
 \begin{itemize}
  \item[i)] The interconnection in Figure \ref{fig:Pfig5} with $\phi(\cdot)$ is well posed.
  \item[ii)] The IQC defined by $\phi( \cdot )$ is satisfied.
  \item[iii)] There exists an $\epsilon$ such that
  \begin{align}
   \begin{bmatrix} G(j \omega) \\ I \end{bmatrix}^{*}
   \Pi (j \omega) 
   \begin{bmatrix} G(j \omega) \\ I \end{bmatrix} < -\epsilon I \quad \forall \omega \in \mathbb{R}
  \end{align}
 Then the interconnection in Figure \ref{fig:Pfig5} is stable.
 \end{itemize}
 \label{thm:IQC}
\end{theorem}

This is the main stability result which will be used in this article: it is a considerable simplification of the original result in \cite{MegretskiIEEETAC1997} and is allowed due to the form of the nonlinearites $\phi(\cdot)$ and the IQC's which will be developed. In general, more attention needs to be paid to the well-posedness of the feedback system and the form of the IQC - see \cite{MegretskiIEEETAC1997,Jonsson:notes}. Much more could be written about IQC's but in an effort to keep the article reasonably brief and readable, further discussion is omitted. An excellent reference on the subject is \cite{Jonsson:notes} and interesting connections between time-domain and frequency domain IQC's are made in \cite{SeilerIEEETAC2016} (see also \cite{PfiferIJRNC2016}). A recent review of IQC's is given in \cite{VeenmanEJC2016}.

\section{Monotone Nonlinearities}

Zames and Falb developed analysis techniques for two important classes of nonlinearity: monotone non-decreasing nonlinearities and slope restricted nonlinearities. In addition, both odd and non-odd versions of these nonlinearities were treated. However, Zames and Falb's results are best appreciated first for monotone nonlinearities, and these results can then be adapted for slope-restricted nonlinearities.
In this section it will be shown how the monotone property can be used to establish an IQC for these classes of nonlinearities.

\subsection{A preliminary result involving convex functions}

Firstly, note that if $\phi \in \mathcal{N}_{M}$, then the integral 
\[
 F(\sigma) = \int_{0}^{\sigma} \phi (x) dx 
\]
is a \emph{convex function}, since its derivative exists and is monotone non-decreasing by assumption, viz
\[
 \frac{d F(\sigma )}{d \sigma} = \phi(\sigma)
\]
Convexity of $F(\sigma)$ implies that, for all $\sigma_1,\sigma_2 \in \mathbb{R}$,
\begin{align}
 & (\sigma_1 -\sigma_2) F'(\sigma_1) \geq F(\sigma_1) - F(\sigma_2) \\
 \Leftrightarrow & 
 (\sigma_1 -\sigma_2) \phi(\sigma_1) \geq F(\sigma_1) - F(\sigma_2)
 \label{eq:convex1}
\end{align}
where $F'(\sigma_1)$ represents the derivative of $F(\sigma)$ at $\sigma_1$. 
Since this inequality holds for any $\sigma_1,\sigma_2 \in \mathbb{R}$, then it holds for $\sigma_1=x(t)$ and $\sigma_2 = x(t - \tau)$, where $t$ and $\tau$ are arbitrary, which gives
\begin{align}
 \big( x(t) -x(t-\tau) \big) \phi(x(t)) \geq F(x(t)) - F(x(t-\tau))
\end{align}
Integrating both sides w.r.t. $t$ then gives
\marginpar{\small \emph{``If, he thought to himself, an infinite improbability machine is a virtual impossibility, then it must logically be a finite improbability.''}}
\begin{align}
 & \int_{-\infty}^{\infty} \big( x(t) -x(t-\tau) \big) \phi(x(t)) dt 
 \geq \underbrace{ \int_{-\infty}^{\infty} \Big( F(x(t)) - F(x(t-\tau)) \Big) dt}_{=0} \\
\Rightarrow & 
\int_{-\infty}^{\infty} \big( x(t) -x(t-\tau) \big) \phi(x(t)) dt 
 \geq 0
\end{align}
which can then be re-arranged to give the remarkable, and perhaps unexpected, result that
\begin{align}
\int_{-\infty}^{\infty} x(t) \phi(x(t)) dt 
 \geq \int_{-\infty}^{\infty} x(t-\tau) \phi(x(t)) dt
 \label{eq:area}
\end{align}
for all $\tau$. Inequality (\ref{eq:area}) is sometimes called the \emph{area inequality}.

In fact, it will also be useful to develop another version of the \emph{area inequality} as well which applies in the special case that $\phi(\cdot)$ is an odd function. 
First note that if $\phi(.)$ is odd then its integral, $F(\cdot)$ is even. Hence replacing $\sigma_2$ with $-\sigma_2$ in inequality (\ref{eq:convex1}) thus gives:
\begin{align}
 (\sigma_1 -(-\sigma_2)) \phi(\sigma_1) & \geq F(\sigma_1) - F(-\sigma_2) \\
 \Leftrightarrow 
 (\sigma_1 +\sigma_2) \phi(\sigma_1) & \geq F(\sigma_1) - F(\sigma_2) 
 \label{eq:convex2}
\end{align}
Again, noting that the above inequality holds for any $\sigma_1,\sigma_2 \in \mathbb{R}$, then it holds for  $\sigma_1=x(t)$ and $\sigma_2 = x(t - \tau)$, which gives
\begin{align}
 \big( x(t) +x(t-\tau) \big) \phi(x(t)) \geq F(x(t)) - F(x(t-\tau))
\end{align}
As before, integrating both sides w.r.t. $t$ then gives
\begin{align}
 & \int_{-\infty}^{\infty} \big( x(t) +x(t-\tau) \big) \phi(x(t)) dt 
 \geq \int_{-\infty}^{\infty} \Big( F(x(t)) - F(x(t-\tau)) \Big) dt \\
\Rightarrow & 
\int_{-\infty}^{\infty} \big( x(t) +x(t-\tau)) \phi(x(t)) \big) dt 
 \geq   0
\end{align}
which can then be re-arranged to give a mirrored result to inequality (\ref{eq:area}):
\begin{align}
\int_{-\infty}^{\infty} x(t) \phi(x(t)) dt 
 \geq - \int_{-\infty}^{\infty} x(t-\tau) \phi(x(t)) dt
 \label{eq:area2a}
\end{align}
Combining inequalities (\ref{eq:area}) and (\ref{eq:area2}) we have
\begin{align}
 - \int_{-\infty}^{\infty} x(t) \phi(x(t)) dt
 \leq \int_{-\infty}^{\infty} x(t-\tau) \phi(x(t)) dt 
 \leq \int_{-\infty}^{\infty} x(t) \phi(x(t)) dt
\end{align}
which of course implies that
\begin{align}
 \left| \int_{-\infty}^{\infty} x(t-\tau) \phi(x(t)) dt \right|
 \leq  \int_{-\infty}^{\infty} x(t) \phi(x(t)) dt 
 \label{eq:area2}
\end{align}
which is an \emph{area inequality} for odd nonlinearities. The two area inequalities can be packaged into the following Lemma.

\newtheorem{lemma}{Lemma}
\begin{lemma}[Area Lemma] 
 Consider the nonlinearity $\phi(\cdot) \in \mathcal{N}_{M}$. Then, for all $\tau \in \mathbb{R}$, inequality (\ref{eq:area}) holds. Furthermore, if $\phi(\cdot) \in \mathcal{N}_{MO}$ then, for all  $\tau \in \mathbb{R}$, inequality (\ref{eq:area2}) holds.
\end{lemma}

This result is covered in \cite{ZamesSIAM1968,Vidyasagar:book,Willems:book} and as Willems notes, has its roots in re-arrangement inequalities \cite{Hardy:book}.

\subsection{Zames-Falb IQC for monotone nonlinearities}
\label{sec:M}

This section establishes an IQC involving a dynamic multiplier $\textbf{M}$, with transfer function $M(s)=1 -H(s)$ and impulse response $m(t)=\delta(t) - h(t)$, for the nonlinearity $\phi \in \mathcal{N}_M$. Consider the integral
\begin{align}
 I_1 = & \langle \mathbf{M} x(t), \phi(x(t)) \rangle \\
     = & \langle (I - \mathbf{H}) x(t), \phi (x(t)) \rangle \\
     = & \langle x(t), \phi(t) \rangle - \langle \mathbf{H} x(t),\phi(x(t)) \rangle \\
     = & \langle x(t), \phi(t) \rangle - \int_{-\infty}^{\infty} \left( \int_{-\infty}^{\infty} h(\tau) x(t-\tau) d \tau \right) \phi(x(t)) dt
\end{align}
Changing the order of integration then gives
\begin{align}
 I_1 = & \langle x(t), \phi(t) \rangle - \int_{-\infty}^{\infty} h( \tau) \left( \int_{-\infty}^{\infty} x(t-\tau) \phi(x(t)) d t \right)  d \tau 
\end{align}
Next, \underline{assuming $h(t) \geq 0 \quad \forall t$}, the area inequality (\ref{eq:area}) means that 
\begin{align}
h(\tau) \int_{-\infty}^{\infty} x(t) \phi(x(t)) dt 
 \geq h(\tau) \int_{-\infty}^{\infty} x(t-\tau) \phi(x(t)) dt
 \label{eq:area2c}
\end{align}
for all $\tau$, which then implies
\begin{align}
 I_1 & \geq \langle x(t), \phi(t) \rangle - \int_{-\infty}^{\infty} h( \tau) \left( \int_{-\infty}^{\infty} x(t) \phi(x(t)) d t \right) d \tau \\
     & = \langle x(t), \phi(t) \rangle - \int_{-\infty}^{\infty} h( \tau) d \tau \int_{-\infty}^{\infty} x(t) \phi(x(t)) d t \\
     & = \langle x(t), \phi(t) \rangle - \| \mathbf{H} \|_1 \langle x(t),\phi(x(t)) \rangle \\
     & = (1-\| \mathbf{H} \|_1) \langle x(t), \phi(t) \rangle
\end{align}
Now since $\langle x(t), \phi(t) \rangle \geq 0$, we see that $I_1 \geq 0$ if $\| \mathbf{H} \|_1 \leq 1$. The results can be summarised in the following theorem.

\begin{theorem}[Monotone Nonlinearities]
 Let $\phi \in \mathcal{N}_{M}$, and let $\mathbf{M} = I -\mathbf{H}$ be such that $h(t) \geq 0$ for all $t \in \mathbb{R}$ and $\| \mathbf{H} \|_1 \leq 1$, then the following time-domain integral quadratic constraint holds:
 \[
  \int_{-\infty}^{\infty} (\mathbf{M} x(t)) \phi(x(t)) \geq 0
 \]
 \label{thm:M}
\end{theorem}

It is more usual to express this in the frequency domain and, indeed using Parseval's Identity, it follows that above ``time-domain'' IQC can be written in the more familiar frequency domain form as
\begin{align}
 \int_{-\infty}^{\infty} \begin{bmatrix}
                          \hat{x} (j \omega)  \\
                          \widehat{\phi(x)}(j \omega)
                         \end{bmatrix}^*
                        \begin{bmatrix} 0 & M^*(j \omega) \\
                         M(j \omega) & 0 
                        \end{bmatrix}
 \begin{bmatrix}
                          \hat{x} (j \omega)  \\
                          \widehat{\phi(x)}(j \omega)
                         \end{bmatrix} d \omega  \geq 0
                         \label{eq:IQC_ZF1}
\end{align}

\subsection{Zames-Falb IQC for monotone odd nonlinearities}
\label{sec:MO}

\marginpar{\small \emph{``Listen, three eyes,'' he said, ``don't you try to outweird me, I get stranger things than you free with my breakfast cereal.''}}

The strange thing about Theorem \ref{thm:M} is that there is a requirement for the impulse response of the dynamic element of the multiplier (that is, $\mathbf{H}$) to be non-negative: $h(t) \geq 0$ for all time. If more is known about the nonlinearity, namely that it is odd and hence $\phi \in \mathcal{N}_{MO}$, the requirement that $h(t) \geq 0$ can be omitted. Consider again the integral:
\begin{align}
 I_1 = & \langle \mathbf{M} x(t), \phi(x(t)) \rangle \\
 & = 
\langle x(t), \phi(t) \rangle - \int_{-\infty}^{\infty} h( \tau) \left( \int_{-\infty}^{\infty} x(t-\tau) \phi(x(t)) d t \right)  d \tau 
\label{eq:ZFo1}
\end{align}
where the same steps have been applied as in the previous section. 
Previously, at this point, positivity of $h(t)$ was invoked, but suppose this is not assumed; instead use will be made of the other version of the area inequality - inequality (\ref{eq:area2}). From equation (\ref{eq:ZFo1}) it follows that
\begin{align}
 I_1 & \geq \langle x(t), \phi(t) \rangle
 - \left|  \int_{-\infty}^{\infty} h( \tau) \left( \int_{-\infty}^{\infty} x(t-\tau) \phi(x(t)) d t \right)  d \tau \right| \\
     & \geq \langle x(t), \phi(t) \rangle -
     \int_{-\infty}^{\infty} |h(\tau)| 
     \left| \int_{-\infty}^{\infty} x(t-\tau) \phi(x(t)) d t \right| d \tau 
\end{align}
Using the second version of the area inequality (\ref{eq:area2}) yields
\begin{align}
 I_1 & \geq \langle x(t), \phi(t) \rangle
 - \int_{-\infty}^{\infty} |h(\tau)| 
     \left( \int_{-\infty}^{\infty} x(t) \phi(x(t)) d t \right) d \tau \\
     & = \langle x(t), \phi(t) \rangle - \| \mathbf{H} \|_1 \langle x(t), \phi(t) \rangle \\
     & = (I-\| \mathbf{H} \|_1) \langle x(t), \phi(t) \rangle
\end{align}
Again, note that if $\| \mathbf{H} \| \leq 1$ it is clear that $I_1 \geq 0$. This result can be stated as the following theorem:

\begin{theorem}[Monotone Odd Nonlinearities]
 Let $\phi \in \mathcal{N}_{MO}$, and let $\mathbf{M} = I -\mathbf{H}$ and be such that $\| \mathbf{H} \|_1 \leq 1$, then the following time-domain integral quadratic constraint holds:
 \[
  \int_{-\infty}^{\infty} (\mathbf{M} x(t)) \phi(x(t)) \geq 0
 \]
 \label{thm:MO}
\end{theorem}

Again, this time-domain integral can be expressed as a more familiar frequency domain IQC (\ref{eq:IQC_ZF1}). 

\section{Slope-restricted nonlinearities} 

Most frequently, the Zames-Falb results are applied to systems where the nonlinearity $\phi(\cdot)$ is a \emph{slope-restricted} nonlinearity: this means that the generalised derivative of the nonlinearity lies between two values. In this work, the lower slope of the nonlinearity is $0$ and the upper slope is some positive number $\alpha$: $\partial \phi \in [0,\alpha]$. A little generality is lost with this, but again, loopshifting can be used to guarantee this; the exposition is also simplified with this assumption.

The trick to applying Zames-Falb multiplier results to systems with slope restrictions is essentially one of re-arranging the system so that instead of working with the original nonlinearity,  a modified nonlinearity is created such that it is simply monotone increasing: this then enables the original results (on monotone nonlinearities) to be applied to the case of slope-restricted nonlinearities. In the original results of Zames and Falb (\cite{ZamesSIAM1968} - see also \cite{Vidyasagar:book}) the transformation of the original nonlinearity into a modified nonlinearity was done using the commonly used absolute stability trick of \emph{loopshifting}. The same approach is effectively used here, but the results are a little more direct. 

\subsection{Zames-Falb IQC for slope-restricted nonlinearities}

 \begin{figure}[!h]
   \includegraphics[width=0.6\textwidth]{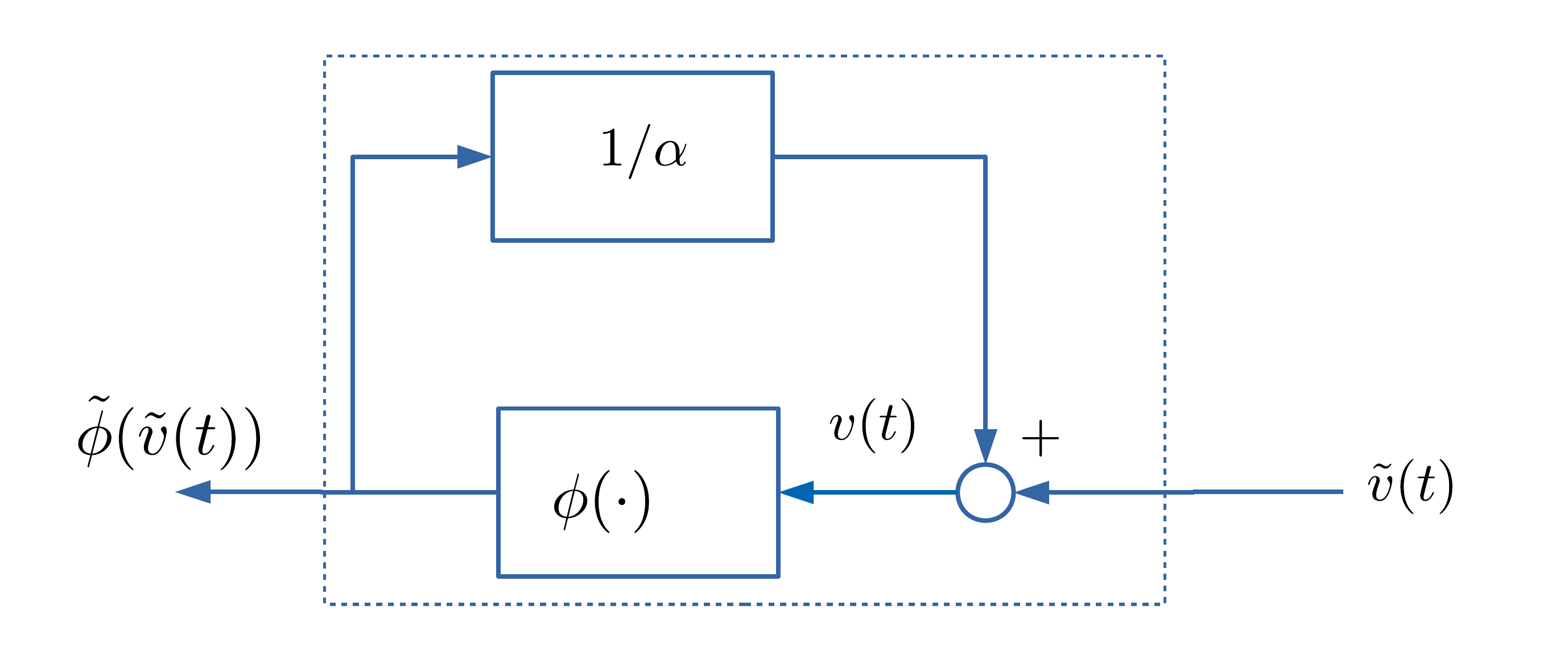}
   \parbox{0.4\textwidth}{ \vspace{-2in}
   \caption{\label{fig:loopshift_non} 
   A loopshifted nonlinearity: $\phi(\cdot)$ is slope-restricted, $\tilde{\phi}(\cdot)$ is monotone. 
   The idea is to establish an IQC for $\tilde{\phi}(\cdot)$ and then use this to find one for $\phi (\cdot)$.
   }}
 \end{figure}

Consider the modified nonlinearity shown in Figure \ref{fig:loopshift_non}. This nonlinearity, $\tilde{\phi}$ can be written as
\begin{align}
  \tilde{\phi}(\tilde{v}) = \left\{
 \begin{array}{ccl}
  \phi(\tilde{v}) & = & \phi(v)\\
  v & = &  \tilde{v} + \frac{1}{\alpha} \phi(v) 
 \end{array} \right.
 \label{eq:phit_implicit}
\end{align}
Assuming solutions exist to this implicit equation, one can write:
\begin{align}
  \tilde{\phi} (\tilde{v}) = \phi \circ (1-\frac{1}{\alpha}\phi(.))^{-1}(\tilde{v})
  \label{eq:phit}
\end{align}
Note that because $\alpha \sigma - \phi(\sigma)$ is a  monotone nonlinearity (by virtue of the slope restriction $\partial \phi \in [0,\alpha]$), then 
\[
 1 - \frac{1}{\alpha} \phi(.)
\]
also is and so is its inverse. Furthermore, again by the slope restriction, $\phi(.)$ is monotonically increasing function, and hence so is $\tilde{\phi}$ defined above. From Section \ref{sec:M}, on monotone nonlinearities, it follows that $\tilde{\phi}$ therefore satisfies the IQC (\ref{eq:IQC_ZF1}) where $\mathbf{H}$ should have a positive impulse response. Note that this IQC essentially defines a relationship between $\tilde{\phi}$ and $\tilde{v}$:  one between $\phi$ and $v$ is actually what is sought. However, from equation (\ref{eq:phit_implicit}), the following relationship holds:
\begin{align}
  \begin{bmatrix}
         v \\ \phi(v) 
        \end{bmatrix}
         = 
        \underbrace{\begin{bmatrix} 
          1 & \frac{1}{\alpha} \\ 0 & 1
         \end{bmatrix}}_{X}
         \begin{bmatrix}
          \tilde{v} \\ \tilde{\phi}(\tilde{v})
         \end{bmatrix}
         \label{eq:relationship}
\end{align}
Therefore if $\tilde{\phi}(.)$ satisfies the IQC defined by $\Pi (j \omega)$ given in equation (\ref{eq:IQC_ZF1}), then the following inequality holds
\begin{align}
 \int_{-\infty}^{\infty}
 \begin{bmatrix}
  \tilde{v}(j \omega) \\ \tilde{\phi}(j \omega) 
 \end{bmatrix}^*
  \begin{bmatrix} 0 & M^* (j \omega) \\ M (j \omega) & 0
    \end{bmatrix}
  \begin{bmatrix}
  \tilde{v}(j \omega) \\ \tilde{\phi}(j \omega) 
 \end{bmatrix} ~ d \omega \geq 0
 \label{eq:IQCmon}
\end{align}
The relationship in equation (\ref{eq:relationship}) can be reversed to get
\begin{align}
  \begin{bmatrix}
         \tilde{v} \\ \tilde{\phi}(\tilde{v}) 
        \end{bmatrix}
         = 
        \underbrace{\begin{bmatrix} 
          1 & -\frac{1}{\alpha} \\ 0 & 1
         \end{bmatrix}}_{X^{-1}}
         \begin{bmatrix}
          v \\ \phi (v)
         \end{bmatrix}
         \label{eq:relationship2}
\end{align}
and then used in inequality (\ref{eq:IQCmon}) to get
\begin{align}
 \int_{-\infty}^{\infty}
 \begin{bmatrix}
  {v}(j \omega) \\ {\phi}(j \omega) 
 \end{bmatrix}^*
 \begin{bmatrix} 
          1 & 0 \\ -\frac{1}{\alpha} & 1
         \end{bmatrix}
  \begin{bmatrix} 0 & M^* (j \omega) \\ M (j \omega) & 0
    \end{bmatrix}
    \begin{bmatrix} 
          1 & -\frac{1}{\alpha} \\ 0 & 1
         \end{bmatrix}
  \begin{bmatrix}
  {v}(j \omega) \\ {\phi}(j \omega) 
 \end{bmatrix} ~ d \omega \geq 0
 \label{eq:IQCmon2}
\end{align}

Therefore $\phi$ satisfies the relationship defined by the matrix
\begin{align}
 {\Pi}_S(j \omega) & = (X^{-1})' \Pi (j \omega) X^{-1} \\
  & = \begin{bmatrix} 1 & 0 \\ -\frac{1}{\alpha} & 1\end{bmatrix}
    \begin{bmatrix} 0 & M^* (j \omega) \\ M (j \omega) & 0
    \end{bmatrix}
    \begin{bmatrix} 
          1 & -\frac{1}{\alpha} \\ 0 & 1
         \end{bmatrix} \\
  & =   \begin{bmatrix} 0 & M^{*}(j \omega) \\
     M(j \omega) & - \frac{1}{\alpha} ( M^{*} (j \omega) + M (j \omega)) 
    \end{bmatrix}
\end{align}
or, multiplying through by $\alpha > 0$, the more familiar form emerges:
\begin{align}
{\Pi}_S (j \omega) = 
 \begin{bmatrix} 0 & \alpha M^{*}(j \omega) \\
     \alpha M(j \omega) & - M^{*} (j \omega) - M (j \omega) 
    \end{bmatrix} \label{eq:IQC_ZF2}
\end{align}

As before, this result can be packaged into the theorem below.

\begin{theorem}[Slope restricted nonlinearities]
 Let $\phi(\cdot)$ be such that $\partial \phi \in [0 , \alpha]$, and let $\mathbf{M} = I -\mathbf{H}$ be such that $h(t) \geq 0$ for all $t$ and $\| \mathbf{H} \|_1 \leq 1$. Then $\phi(\cdot)$ satisfies the IQC defined by ${\Pi}_S(j \omega)$ in inequality (\ref{eq:IQC_ZF2}). 
\end{theorem}

\subsection{Zames-Falb IQC for odd slope-restricted nonlinearities} 

The previous subsection effectively explained how a slope-restricted nonlinearity could be transformed into a monotone nonlinearity and how the results of Section \ref{sec:M}, or more precisely Theorem \ref{thm:M}, could be used to derive an IQC for the original slope restricted nonlinearity. The same is done in this section, but the additional assumption that $\phi(\cdot)$ is odd will be imposed: this then allows the results of Section \ref{sec:MO} to be used, and in particular Theorem \ref{thm:MO}, to be harnessed to provide a similar result. 

Firstly, again consider the modified nonlinearity depicted in Figure \ref{fig:loopshift_non}. This can again be described by the implicit equation (\ref{eq:phit_implicit}) or, more compactly by equation (\ref{eq:phit}). Note that if it is assumed that $\phi(.)$ is odd, then this implies that $\tilde{\phi}(\cdot)$ is also odd because the composition and inverse of odd functions is also odd. Therefore, it follows that $\tilde{\phi}(\cdot)$ is both odd and monotone (from the discussion in the previous section). With this in mind, from Theorem \ref{thm:MO}, $\tilde{\phi}(.)$ satisfies the IQC defined by (\ref{eq:IQC_ZF1}). In exactly the same way as in the previous subsection, using the relationship (\ref{eq:relationship}), it thus follows that $\phi(\cdot)$ satisfies the IQC defined by (\ref{eq:IQC_ZF2}) \emph{except} without the stipulation that the impulse response of the dynamic part of the multiplier, $\mathbf{H}$, is positive. The conclusion of this section can be summarised in the following theorem.

\begin{theorem}[Odd Slope restricted nonlinearities]
 Let $\partial \phi \in [0 , \alpha]$ be an odd nonlinearity, and let $\mathbf{M} = I -\mathbf{H}$ with $\| \mathbf{H} \|_1 \leq 1$. Then $\phi(.)$ satisfies the IQC defined by ${\Pi}_S(j \omega)$ in inequality (\ref{eq:IQC_ZF2}). 
\end{theorem}

\subsection{Comparison of Results}

\begin{figure}[!h]
 \centering
 \includegraphics[width=0.8\textwidth,trim={.5in .7in .5in .9in},clip]{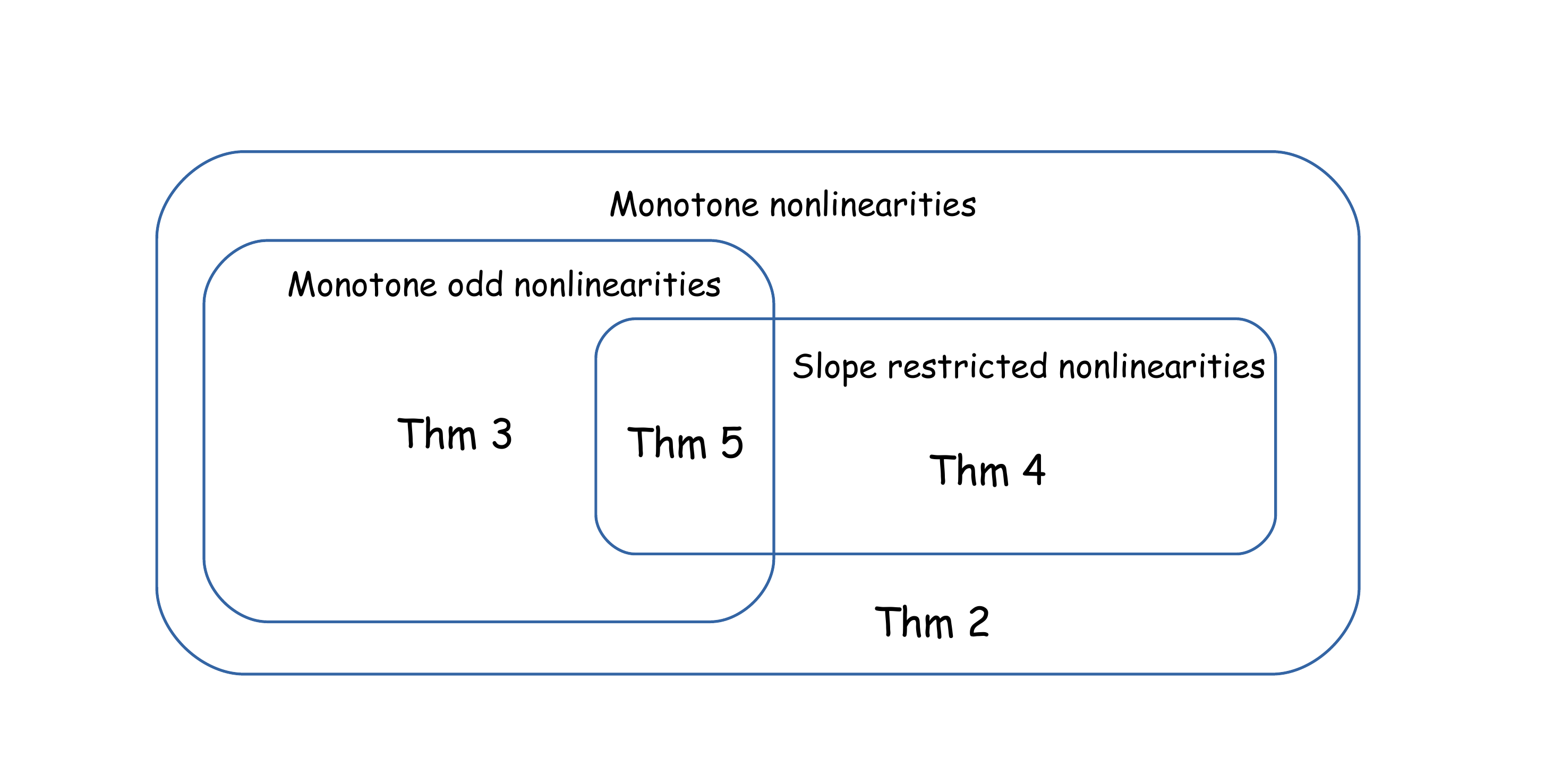}
 \caption{\label{fig:venn} Venn diagram showing relationship between the various Zames-Falb results, assuming the lower bound on the slope is zero.} 
\end{figure}

The relationship between the results is perhaps best summarised by the Venn diagram in Figure \ref{fig:venn}, assuming that the lower slope restriction is zero. Note that Theorem \ref{thm:MO} is a special case of Theorem \ref{thm:M} with the added restriction that $\phi(\cdot)$ should be odd, but with the corresponding relaxation that the impulse response of $\mathbf{H}$ need not be positive. Likewise Theorem 4 is a special case of Theorem 3 but with the added restriction that $\phi(\cdot)$ should be odd, but again with the relaxation that the impulse response of $\mathbf{H}$ need not be positive.

\section{Notes and Further Reading}

The purpose of these notes was to give a clear exposition of the theory behind Zames-Falb multipliers. The subject is still an active area of research and for more up-to-date results, one is advised to consult the latest literature in this area. In this section a few brief additional comments are given but the comments are indeed brief and are by no means self-contained. 

\subsection{Applying Zames-Falb multipliers - searches} 

\marginpar{\small \emph{Infinity is just so
big that by comparison, bigness itself looks really titchy.
Gigantic multiplied by colossal multiplied by staggeringly
huge is the sort of concept we're trying to get across here.}}

Using Zames-Falb multipliers for stability analysis is significantly more difficult than using the Circle or Popov criteria because one has to somehow \emph{find or choose} the dynamics of the Zames-Falb multiplier, $\mathbf{M}$. This is made more difficult by the observation that they may be drawn from a \emph{very wide range of linear operators}: it was noted in the original work of Zames and Falb that these operators do not have to even be rational or causal \footnote{Irrationality has been exploited in the work of \cite{GapskiIEEETAC1994,ChangIEEETAC2011,CarrascoIEEETAC2020}.}. The natural question is then: how does one pick ``the best'' Zames-Falb multiplier for stability analysis? Or more precisely: what is the Zames-Falb multiplier associated with the largest slope restriction or the smallest \Ltwo gain (of an appropriate mapping)?

In the frequency domain, for slope-restricted nonlinearities, the IQC results of Theorem \ref{thm:IQC} mean that the following inequality must be verified for some $\epsilon > 0$:
\begin{align}
 \begin{bmatrix}
  G(j \omega) \\ I 
 \end{bmatrix}^*
  \begin{bmatrix}
   0 & \alpha M(j \omega)^* \\ \alpha M(j \omega) & - M(j \omega) - M^* (j \omega)
  \end{bmatrix}
 \begin{bmatrix}
  G(j \omega) \\ I 
 \end{bmatrix} \leq -\epsilon \\
 M(j \omega)^* (\alpha G(j \omega)^* -I ) +
 (\alpha G(j \omega)-I)M(j\omega) \leq -\epsilon
 \label{eq:passivity}
\end{align}
To assist in the search, the above inequality is often transformed into a search over symmetric matrices using the KYP Lemma (see \cite{RantzerSCL1996}). This yields a matrix inequality of the form
\begin{align}
 \begin{bmatrix}
  A' \mathbf{P} + \mathbf{P} A & \mathbf{P} B + C' 
  \\ 
  B' \mathbf{P} + C & D + D' 
 \end{bmatrix} < 0 \quad \mathbf{P} = \mathbf{P}'
\end{align}
Unfortunately, matrix inequality is bilinear because the state-space matrices $(A,B,C,D)$ are affine functions of the multiplier parameters.  Therefore, one still needs to find some way in to search, efficiently, for an appropriate multiplier. It is important to emphasize that:

\begin{center}
\parbox{0.8\textwidth}{\centering
\emph{The \underline{frequency domain} ``passivity'' constraint (\ref{eq:passivity}) must also be combined with a \underline{time-domain} constraint enforcing $\| \mathbf{H} \|_1 \leq 1$  (sometimes called the $\mathcal{L}_1$ condition).}}
\end{center}
This observation is, effectively, what makes analysis using Zames-Falb multipliers challenging in practice. 

There are three prevailing approaches to Zames-Falb multiplier searches in the literature. Perhaps the most common was proposed by \cite{Chen95}  which, in essence, requires one to choose multipliers of a particular form:
\begin{align}
 M(s) = 1 - H(s) = 1 - \prod_{i=1}^n \frac{k_i}{s+a_i}
\end{align}
Note that in the case that $\mathbf{M} \in \mathcal{RH}_{\infty}$ $a_i > 0$ and therefore choosing $k_i > 0$ one immediately has that $h(t) \geq 0 $ for all $t$. The $\mathcal{L}_1$ condition then simply becomes a case of ensuring $\| \mathbf{H} \|_1 = \prod_{i=1}^n k_i/a_i \leq 1$. Equipped with this form, one can then simply insert this multiplier in equation (\ref{eq:IQC_ZF2}) and use this to evaluate the frequency domain inequality in Theorem \ref{thm:IQC}. Note that this frequency domain inequality is equivalent to a matrix inequality by the KYP Lemma \cite{RantzerSCL1996}. The main issue with this approach is that one needs to choose the poles of the multiplier $a_i$ to ensure that the arising matrix inequalities are linear. The key advantage is that one has an exact expression for the $\mathcal{L}_1$ norm of $\mathbf{H}$. References \cite{SchererIJRNC2014,VeenmanEJC2016} discuss this approach further.

\marginpar{\small \emph{...man had always assumed that he was more intelligent than dolphins because he had achieved so much - the wheel, New York, wars and so on - whilst all the dolphins had ever done was muck about in the water having a good time. But conversely, the dolphins had always believed that they were far more intelligent than man - for precisely the same reasons.}}

An alternative approach to the above, which was originally suggested in \cite{TurnerIEEETAC2009} and then used in \cite{CarrascoSCL2014,CarrascoIEEETAC2020,TurnerIEEETAC2020,TurnerIJRNC2021} and elsewhere, is to assume a particular order of multiplier (at least plant-order) and then to use a change of variables similar to that used in \cite{SchererIEEETAC1997} to obtain convex conditions for stability. The approach is algebraically involved and either uses a bound on the $\mathcal{L}_1$ norm \cite{SchererIEEETAC1997}, which is known to be extremely conservative in some cases; or one can use symmetric multipliers \cite{TurnerIEEETAC2020} which remove the conservatism in the $\mathcal{L}_1$ norm, but then one must accept the phase restrictions imposed by the symmetry of the multiplier. My own - and somewhat biased view - is that this general approach is somewhat more intellectually satisfying than the previous one where effectively one makes, in practice, some arbitrary choices for poles and order - but improvements need to be made to overcome the current conservatism which plagues the approach in some cases.

The final approach that is worth mentioning is that in Zames and Falb's original results, the multiplier $\mathbf{M}$ may actually be chosen \emph{irrational} and this has been exploited by several authors \cite{GapskiIEEETAC1994} and \cite{ChangIEEETAC2011} in order to obtain stability results. Recently \cite{CarrascoIEEETAC2020} Carrasco, Heath and co-authors have proposed  discrete-time search for multipliers, based on FIR functions, which when interpreted in continuous time provides irrational multipliers which appear to be competitive with the state of the art. The potential disadvantage with this technique is that it may not be easy to combine this with synthesis techniques. 

\begin{table}[!t]
\centering
 \begin{tabular}{|l|l|l|l|}
 \hline 
  Approach & Order & $\mathcal{L}_1$ bound & Comments \\ \hline
  \cite{Chen95,SchererIJRNC2018} & chosen & exact & User must choose order and poles \\
  \cite{TurnerIEEETAC2009,CarrascoSCL2014} &
  plant order & upper bound & Line search required \\
  \cite{TurnerIEEETAC2020} &
  2$\times$ plant order & exact & Non-causal multipliers; symmetric state-space structure \\
  \cite{ChangIEEETAC2011} & irrational & exact & Competitive but perhaps fragile \\ \hline
 \end{tabular}
 \caption{\label{tab:1} Comparison between some different approaches for proving stability with Zames-Falb multipliers}
\end{table}

Table \ref{tab:1} gives a rough comparison between the three different approaches: a detailed comparison is out of scope and the interested reader should consult \cite{CarrascoEJC2016} or \cite{SchererIJRNC2018} for more detailed discussion. It is interesting that opinions on the superiority of a particular form of multiplier vary and that the same property may be perceived to be an advantage by some researchers, but a disadvantage by others. At this point it is fair to say ``the jury is out'' on which approach is best. 

\subsection{Multivariable systems}

This article, and  the original results of Zames and Falb, has been dedicated to the case when $\phi(\cdot)$ is a scalar nonlinearity. More generally, $\phi (\cdot): \mathbb{R}^m \mapsto \mathbb{R}^m$ and hence multivariable generalisations of Zames-Falb multipliers must be used. This extension was only carried out relatively recently by Safonov and co-authors in a number of papers \cite{SafonovIJRNC2000,SafonovIEEETAC2002,SafonovSCL2005}. The generalisations are not trivial and require further assumptions on the form of the nonlinearity to be made, but they do have IQC interpretations in much the same way as their scalar-valued counterparts \cite{Amato2001}. The search for multipliers is somewhat more involved however: the generalisation of \cite{TurnerIEEETAC2009} to the multivariable case \cite{TurnerIJRNC2015} is effectively too conservative and too computationally demanding\footnote{The word ``tractable'' in the title is, in hindsight, misleading; ``intractable'' might be more appropriate.}. The approaches given in \cite{SchererIJRNC2018,TurnerCDC2019} both have merit but require choices to be made which are not entirely systematic. Therefore, while it seems that much of the spirit of the Zames-Falb multipliers carries over to the multivariable case, the search for suitable multipliers seems inherently more difficult \footnote{This section needs improvement and more detail. Similar sentiments probably apply to the article as a whole.}.

\subsection{How conservative are Zames-Falb multipliers?}

It has been known for some time that Zames-Falb multipliers are potentially much less conservative than either the Circle or Popov criteria -- in the sense that they may conclude stability when the Circle/Popov Criteria fail to do so. It is not yet certain whether a Zames-Falb Multiplier is a \emph{necessary} as well as \emph{sufficient} condition for absolute stability. Results in both continuous (\cite{Suarxiv2020}) and discrete time (\cite{Carrascoarxiv2020}) seem to suggest that the existence of a Zames-Falb multiplier is fairly close to necessary\footnote{It is believed by some that W.P. Heath's dog has worked out the necessary conditions.} for absolute stability, and thus using Zames-Falb multipliers should give the control engineer some confidence in the results obtained.

\subsection{A historical note}

An often overlooked contributor to the development of Zames-Falb multipliers is the person at their origin: R O'Shea \footnote{It was J. Carrasco and W.P. Heath who impressed upon me the importance of R. O'Shea in the development of Zames-Falb multipliers. His contribution is acknowledged in the paper \cite{CarrascoEJC2016}.}. In a few papers \cite{OShea66,Oshea67}, O'Shea introduces the idea of the sort of multipliers which have now become known as ``Zames-Falb'' multipliers. Although O'Shea did not quite work out the precise mathematics of the multipliers, he should at least be credited with their inception.

\section{Conclusion} 

\marginpar{\small \emph{You live and learn. At any rate, you live.}}

The purpose of this article was to introduce the main technical ideas behind Zames-Falb multiplier's in a reasonably accessible way. After finalising the article, it was evident that I skimmed over some topics and, in places, introduced notation which was either excessive, unclear or both. These two deficiencies have undoubtedly jeopardised my original intention of readability. Then, of course, there are the inevitable typos which will have exacerbated this. This document is, however, something of a working document which I intend to update; in the next version I hope to address some of the weaknesses of this version. 

\section{Acknowledgements}

I have been studying Zames-Falb multipliers for over 10 years and during that time I have greatly benefitted from discussions on the subject with Drs. J. Carrasco, R. Drummond, J.M. Gomes da Silva Jr., W.P. Heath, M.L. Kerr, and J. Sofrony, and also some of the students who took my Nonlinear Control course and other colleagues at various conferences. I'd like to also thank Dr. A.L.J. Bertolin for pointing out some glaring inconsitencies in Section 4.1 of a previous version which have now (hopefully) been corrected. Her work on controller design using Zames-Falb multipliers is of independent interest \cite{BertolinIFAC2022}.

\bibliographystyle{plain}
\bibliography{ZFM_DP_v2a}

\end{document}